\documentclass[aps,prb,twocolumn,amsmath,amssymb,longbibliography]{revtex4-1}

\usepackage[utf8]{inputenc}
\usepackage{graphicx}
\usepackage{booktabs}
\usepackage{notes2bib}
\usepackage{amsmath}
\usepackage[squaren]{SIunits}

\usepackage[colorlinks=true,bookmarks=false,citecolor=blue,urlcolor=blue,pdfstartview=FitH]{hyperref}

\DeclareMathOperator*{\argmax}{argmax}

\begin{document}

\title{Quasinormal mode representation of radiating resonators in open phononic systems}
\author{Vincent Laude}
\affiliation{Institut FEMTO-ST, CNRS UMR 6174, Université Bourgogne Franche-Comté, Besançon, France}
\author{Yan-Feng Wang}
\affiliation{Department of Mechanics, School of Mechanical Engineering, Tianjin University, 300350 Tianjin, China}

\begin{abstract}
Open phononic systems including resonators radiating inside an unbounded medium support localized phonons characterized by a complex frequency.
In this context, the concept of elastic quasinormal mode (QNM) arises naturally, as in the cases of nanophotonic and plasmonic open systems.
Based on a complex, unconjugated form of reciprocity theorem for elastodynamics, the eigenfunction expansion theorem expressed on the elastic QNM basis yields an accurate approximation to the response function, for an arbitrary excitation.
The description of the elastic Purcell effect then requires defining a complex-valued modal volume for each QNM.
For validation, we first consider the case a vibrating nylon rod radiating in water.
As a second test example, we consider a slender nickel ridge on the surface of a fused silica substrate, before extending our attention to a nanoscale tuning fork composed of two such ridges.
In all cases, the response estimated from only a few elastic QNMs agrees with the solution to the elastodynamic equation.
\end{abstract}

\maketitle

\section{Introduction}
\label{sec1}

Resonating elements are ubiquitous in systems supporting wave propagation, including photonics, plasmonics, acoustics and phononics.
The description of wave radiation from discrete resonators in interaction with an open, infinite substrate or a surrounding medium is thus an important problem.
Fundamental questions in this regard include the description of radiation loss from a resonant source, and the enhancement by its environment of the spontaneous emission rate from a quantum system, known as Purcell effect \cite{purcellPR1969}, as well as its extension to classical waves \cite{sauvanPRL2013,yanPRL2020,landiPRL2018,schmidtPRL2018,huangFR2022}.
In photonics and plasmonics, the radiation from resonating objects whose dimensions are smaller than the wavelength is indeed strongly influenced by the surrounding media.
Here, we consider the similar case of small resonators coupled to elastic waves, or acoustic phonons in the long wavelength limit.
In phononics, arrangements of small size resonators on a surface are often considered, either as crystal arrays \cite{achaouiPRB2011}, acoustic metamaterials \cite{colombiSR2016}, or simply systems of a few coupled resonators at the micro- and nanoscale \cite{raguinNC2019}.

An important issue with open systems is the definition and proper use of complex eigenmodes satisfying radiation boundary conditions at infinity.
An open system is not conservative because energy can escape it.
As a result, the dynamical matrix describing wave propagation is not Hermitian and the eigenfunctions are no longer normal modes but quasinormal modes (QNMs) whose frequencies are complex \cite{chingRMP1998}.
Quasinormal mode analysis allows one to rely on only a few QNMs to provide an approximate description of the response, even though they do not respect the orthogonality property of normal modes.
QNMs are used in the description of gravitational waves emitted by perturbed black holes or relativistic stars \cite{nollertCQG1999,bertiPRD2004,torresPRL2020}.
They are widely employed in photonics \cite{benzaouiaPRR2021} and plasmonics \cite{dezfouliO2017} as a practical reduced-order (few-parameter) model based on the resonant frequencies.

In this paper, following ideas from Ref. \cite{sauvanPRL2013}, we elaborate on the concept of elastic quasinormal mode and its application in phononics.
Though there have been previous attempts at defining elastic QNMs based on Green's functions techniques \cite{elsayedPRR2020}, we instead derive our results from basic solutions of the elastodynamic wave equation and a complex, unconjugated form of the reciprocity theorem valid for open systems.
This approach significantly avoids reference to an energy conservation principle.
Furthermore, we have used the concept of the perfectly matched layer (PML) to approximate radiation at infinity and thus obtain QNMs of resonators with arbitrary shape.
Of particular relevance is the use of the superposition of a few QNMs to predict the elastodynamic response to an arbitrary excitation of a resonator.
In the process, we define a complex modal volume and give an expression of the response near resonance that is similar to Purcell's.

\section{Quasinormal mode expansion}
\label{sec2}

\subsection{Normal modes}
\label{sec2A}

In this subsection, we summarize some important properties of the modes of closed and lossless elastodynamic systems, that belong to the class of normal modes.
The purpose is mainly to highlight which properties are not conserved in open systems.

Normal modes are eigenmodes of \textit{closed} structures.
Mathematically, for elastic waves they are the eigensolutions inside a finite domain $\Omega$ (see Figure \ref{fig1}(a)) of the elastodynamic equation
\begin{align}
\omega_n^2 \rho \mathbf{u}_n &= - \nabla \cdot (c : S_n), \label{eq1} \\
S_n &= \nabla \mathbf{u}_n,
\end{align}
with exterior boundary conditions on $\partial\Omega$ (typically free or clamped).
The elastic tensor $c$ has four indices and is symmetrical, $\rho$ is the mass density, $\mathbf{u}$ is the displacement field, and $S$ is the strain tensor.
In the absence of loss, eigenfrequencies $\omega_n$ are real and eigenvectors $\mathbf{u}_n$ are orthogonal.
By projection on normal mode number $m$, the orthogonality relation can be written
\begin{align}
\omega_n^2 \int_\Omega \mathbf{u}_m^* \cdot \rho \mathbf{u}_n &= \int_\Omega S_m^* : c : S_n = 0 \textrm{ if } m \neq n. \label{eq3}
\end{align}
For $m=n$, the equality of kinetic and elastic energy of the normal mode is
\begin{align}
\omega_n^2 \int_\Omega \mathbf{u}_n^* \cdot \rho \mathbf{u}_n &= \int_\Omega S_n^* : c : S_n. \label{eq4}
\end{align}
The total energy of normal modes is bounded
\begin{align}
H(\mathbf{u}_n) &= \frac{1}{2} \left( \int_\Omega S_n^* : c : S_n + \omega_n^2 \int_\Omega \mathbf{u}_n^* \cdot \rho \mathbf{u}_n \right) \nonumber \\
&= \omega_n^2 \int_\Omega \mathbf{u}_n^* \cdot \rho \mathbf{u}_n < \infty. \label{eq5}
\end{align}

If normal modes are known, the eigenexpansion theorem states that any solution $\mathbf{u}$ to the elastodynamic equation at frequency $\omega$
\begin{align}
- \nabla \cdot (c : \nabla \mathbf{u}) - \omega^2 \rho \mathbf{u} &= \mathbf{F} \label{eq6}
\end{align}
can be written $\mathbf{u}(\omega) = \sum_m \alpha_m(\omega) \mathbf{u}_m$ with the frequency-dependent coefficients $\alpha_m(\omega)$.
Combining the equations above, especially the orthogonality relation, it is easy to see that
\begin{align}
\mathbf{u}(\omega) &= \sum_m \frac{1}{\omega_m^2 - \omega^2} \frac{\int_\Omega \mathbf{u}_m^* \cdot \mathbf{F}}{\int_\Omega \mathbf{u}_m^* \cdot \rho \mathbf{u}_m} \mathbf{u}_m. \label{eq7}
\end{align}
The formula thus expresses the response of the elastic system to any excitation $\mathbf{F}$, simply from its projection on each normal mode and the superposition of poles centered on the real eigenfrequencies.

\subsection{Elastic quasinormal modes}
\label{sec2B}

\begin{figure}[tb]
\centering
\includegraphics[width=85mm]{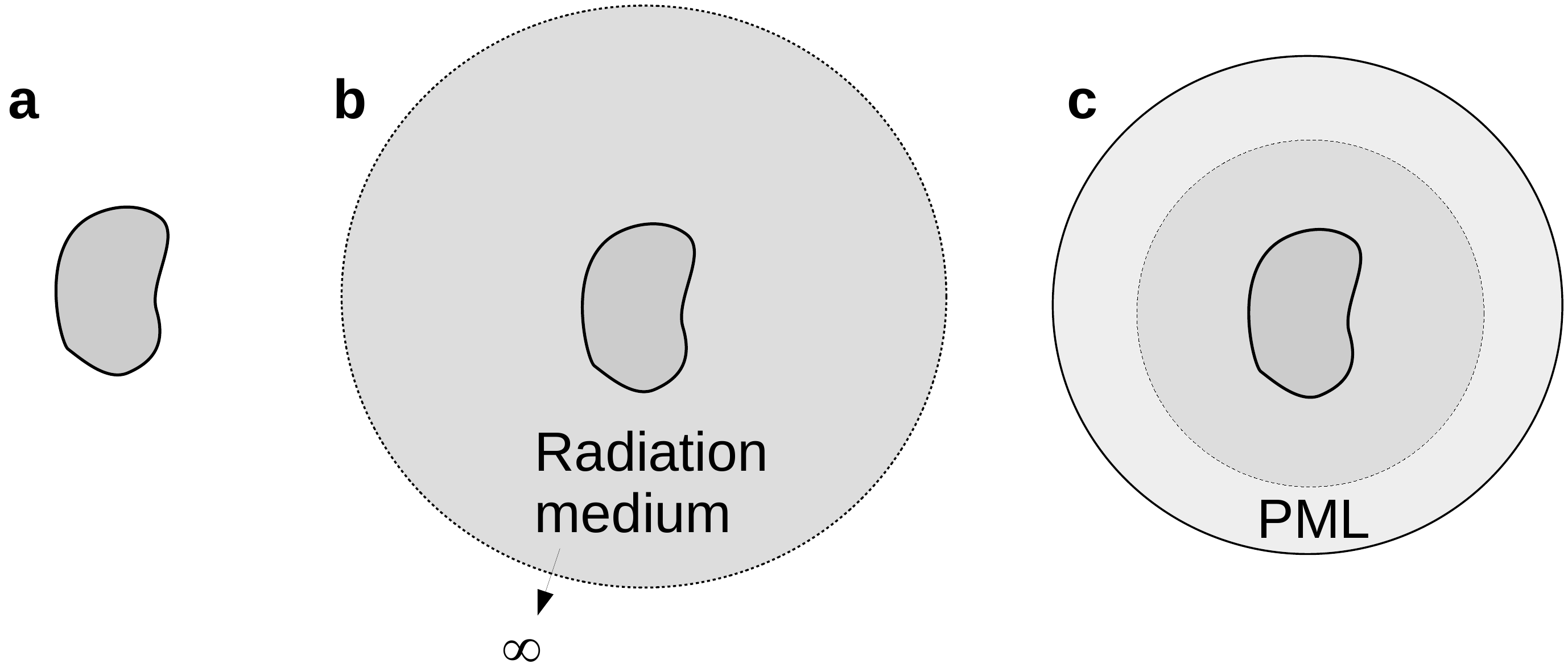}
\caption{Definition of supporting domains for wave resonance and propagation. (\textbf{a}) Finite, closed domain supports normal modes. (\textbf{b}) Infinite, open domain supports quasinormal modes. (\textbf{c}) These can be approximated by closing the domain of computation with a perfectly matched layer (PML), that is the truncated image of the infinite domain in (\textbf{b}) in a complex coordinate transformation.}
\label{fig1}
\end{figure}

As noted in introduction, quasinormal modes are a generalization of normal modes for \textit{open} and lossy systems.
The basic equation \eqref{eq1} defining them is the same, but $c$ and $\rho$ are now complex valued, and possibly dispersive, $\Omega$ is an infinite domain (see Figure \ref{fig1}(b)), and outgoing-wave boundary conditions are considered at infinity.
As a result, the orthogonality relation \eqref{eq3} is lost, as are the finiteness of the total energy \eqref{eq5} and the eigen-expansion of equation \eqref{eq7}.
All eigenfrequencies are now complex valued, since the matrices involved are not symmetric anymore.
As we will show in the next subsection, an expansion over elastic quasinormal modes formula replacing \eqref{eq7} can be obtained anyway.

There are different ways to obtain QNMs in practice.
A rigorous way is to use Green's function techniques \cite{doostPRA2014}, for instance based on some approximation in the finite region (e.g. finite element analysis), coupled to an exterior analytic solution when it is known.
They can be plane waves for planar geometries, Bessel and Hankel functions for cylindrical geometries, or spheroidal harmonics in three-dimensional homogeneous space.
Since we consider vector elastic waves in anisotropic elastic media, such an approach is cumbersome.
Fortunately, there is an efficient way to circumvent the issue, though it is an approximate solution.
That solution is to approximate the infinite radiation medium with a finite perfectly matched layer (PML).
PML is here implemented as a coordinate transformation in the complex plane.
Some eigenvalue solver can then be used to obtain the complex eigenmodes of the now 'closed' system (see Figure \ref{fig1}(c)) but with complex-valued material constants.
A single finite element mesh can be prepared to describe the resonator and the radiation medium.
However, this approach is not always as easy as it seems, since perfectly matched layers have their own eigenmodes, from which the QNMs have to be sorted.

As a note, in the previous subsection on normal modes we have written the integrals rather casually, without mention of the integration variable.
For quasinormal modes, because of the presence of the PML, the domain of integration $\Omega$ is still finite but obtained from a complex coordinate transformation from an infinite domain; the coordinate transformation is characterized by a Jacobian matrix $J$ that is itself a function of spatial coordinates.
The weak form representation of Eq. \eqref{eq1} is
\begin{align}
\omega_n^2 \int_\Omega \mathbf{v} \cdot \rho \mathbf{u}_n |J| d\mathbf{r} &= \int_\Omega S(\mathbf{v}) : c : S(\mathbf{u}_n) |J| d\mathbf{r} \label{eq8}
\end{align}
with $\mathbf{v}$ a test function and with the strain defined as $S(\mathbf{u}) = J^{-t} \nabla \mathbf{u}$.
Note the absence of complex conjugation compared to the normal mode case of Eq. \eqref{eq4}.
The expression for the Jacobian $J$ depends on the PML form that is chosen; in this work we have used the polynomial PML model discussed in Ref. \cite{skeltonWM2007}.
As a note, $J$ is a function of space coordinates but also of frequency.

In practice, we have used in this work the following algorithm to obtain one QNM at a time, as inspired by the power iteration technique \cite{gubernatisJCP2008}.
We start with a guess for the eigenfrequency $\omega_0$ that is close to a maximum of the frequency response.
Formally, we assume that a stiffness matrix $K$ and a mass matrix $M$ have been prepared from Eq. \eqref{eq8}.
The initialization of the algorithm is stochastic: solve $(K - \omega_0^2 M) \mathbf{u}_{0} = \mathbf{F}$ for a random excitation $\mathbf{F}$.
Then the linear problem $(K - \omega_n^2 M) \mathbf{u}_{n+1} = M \mathbf{u}_{n}$ is iteratively solved and the solution converges to the nearest eigenvector.
At the end of the $n$-th iteration, the eigenvector is normalized by its infinite norm $|\mathbf{u}_{n}|_\infty$.
The next candidate eigenfrequency is evaluated as $\omega_{n+1}^2 = \mathbf{u}_{n} \cdot K \cdot \mathbf{u}_{n} / \mathbf{u}_{n} \cdot M \cdot \mathbf{u}_{n}$.
Convergence of this power algorithm is very fast but the solution depends acutely on the distance in the complex plane between the initial frequency and the target QNM frequency.
Note the method is compatible with dispersive media (resulting from the presence of the PML), i.e. $K$ and $M$ are simply updated at the $n$-th iteration as $K(\omega_n)$ and $M(\omega_n)$.

In the following subsection we assume that all necessary QNMs have been obtained and that they form a complete basis for representing solutions to the elastodynamic problem.
Examples of QNMs obtained with the algorithm of this subsection are given in Section \ref{sec3} for two representative examples.

\subsection{Sauvan's method transposed to elastic waves}
\label{sec2C}

Sauvan et al. \cite{sauvanPRL2013} base their derivation of the QNM expansion on a particular form of the electromagnetic reciprocity theorem, the unconjugated form expressed for two arbitrary solutions at different frequencies.
We will keep the latter idea but work directly with the elastic equations of motion; the usual form of the reciprocity theorem for elastic waves is recalled for completeness in appendix A.
We consider the weak form of the equations of motion for a solution $\mathbf{u}_1$ at frequency $\omega_1$ with the test function chosen as another solution $\mathbf{u}_2$ at frequency $\omega_2$, i.e.
\begin{align}
\int S_2 : c(\omega_1) : S_1 - \omega_1^2 \int \mathbf{u}_2 \cdot \rho(\omega_1) \mathbf{u}_1 &= \int \mathbf{u}_2 \cdot \mathbf{F}_1
\label{eq9}
\end{align}
and the same equation with indices $1$ and $2$ permuted.
Their difference then leads to
\begin{align}
& \int S_2 : [c(\omega_1) - c(\omega_2)] : S_1 \nonumber \\
& ~~~~~ - \int \mathbf{u}_2 \cdot [\omega_1^2 \rho(\omega_1) - \omega_2^2 \rho(\omega_2) ] \mathbf{u}_1 \nonumber \\
& ~~~~~ = \int \mathbf{u}_2 \cdot \mathbf{F}_1 - \mathbf{u}_1 \cdot \mathbf{F}_2.
\label{eq10}
\end{align}
This is a reciprocity relation without complex conjugation, valid for an arbitrary frequency-dependent material distribution.
Note that the integration variable is not written explicitly in this section, for compactness of expressions, but all integrals have an implied $|J| d\mathbf{r}$ factor as in Eq. \eqref{eq8}.
We also use the notation $S_n = S(\mathbf{u}_n)$ for the strain tensor.

Next we take solution $2$ as QNM number $n$ and solution $1$ as the current solution $\mathbf{u}$ depending on $\omega$ as a continuous parameter, such that
\begin{align}
& \int S_n : [c(\omega) - c(\omega_n)] : S(\mathbf{u}) \nonumber \\
& ~~~~~ - \int \mathbf{u}_n \cdot [\omega^2 \rho(\omega) - \omega_n^2 \rho(\omega_n) ] \mathbf{u} \nonumber \\
& ~~~~~ = \int \mathbf{u}_n \cdot \mathbf{F}, \forall n .
\label{eq11}
\end{align}
The QNMs constitute a basis for the solution (per the eigenfunction expansion theorem), according to which we can write
\begin{align}
\mathbf{u}(\omega) &= \sum_m \alpha_m(\omega) \mathbf{u}_m .
\label{eq12}
\end{align}
Inserting the eigenfunction decomposition we obtain
\begin{align}
\label{eq13}
\sum_m B_{nm}(\omega) \alpha_m(\omega) = \int \mathbf{u}_n \cdot \mathbf{F} = F_n, \forall n
\end{align}
with
\begin{align}
\label{eq14}
B_{nm}(\omega) &= \int S_n : [c(\omega) - c(\omega_n)] : S_m \nonumber \\
&- \int \mathbf{u}_n \cdot [\omega^2 \rho(\omega) - \omega_n^2 \rho(\omega_n) ] \mathbf{u}_m .
\end{align}
If the QNMs are known, the $B_{nm}(\omega)$ coefficients are easily computed, and the $\alpha_m(\omega)$ are obtained by solving a small linear problem as a function of frequency, formally $\mathbf{\alpha}(\omega) = B(\omega)^{-1} \mathbf{F}$.
By small, we mean that the size of the problem depends on the number of quasinormal modes that are used in practice in the expansion.
It is clear that $B_{nm}(\omega_n) = 0$ by construction.
Applying the reciprocity relation \eqref{eq10} with $\omega_1 = \omega_m$ and $\omega_2 = \omega_n$, we also have $B_{nm}(\omega_m) = 0$ for $m \neq n$.
For all other frequencies, however, $B_{nm}(\omega)$ has in principle a non vanishing value that must be taken into account in the solution.
It is then apparent that matrix $B(\omega)$ is singular at each QNM, in the complex plane, but is always invertible for $\omega$ taken along the real axis.
Finally, equation \eqref{eq12} gives the general solution, i.e. the frequency response of the system to an arbitrary body force distribution.

As a note, if the material constants are non dispersive, the formulas is simplified as
\begin{align}
B_{nm}(\omega) &= (\omega_n^2 - \omega^2) \int \mathbf{u}_n \cdot \rho \mathbf{u}_m.
\label{eq15}
\end{align}
Anyhow, the orthogonality relation of normal modes does not apply and matrix $B(\omega)$ is not diagonal.
The explicit expansion \eqref{eq6} still does not apply.

\bigskip
More can be said regarding the form of the solution close to a resonance, that is in the vicinity of a particular $\omega_n$.
Sauvan's trick \cite{sauvanPRL2013} for this purpose is indeed to pole Eq. \eqref{eq13} by defining
\begin{align}
A_{nm}(\omega) &= \frac{1}{\omega - \omega_m} B_{nm}(\omega) \nonumber \\
&= \frac{1}{\omega - \omega_m} \left[ \int S_n : [c(\omega) - c(\omega_n)] : S_m \right. \nonumber \\
&- \left. \int \mathbf{u}_n \cdot [\omega^2 \rho(\omega) - \omega_n^2 \rho(\omega_n) ] \mathbf{u}_m \right] .
\label{eq16}
\end{align}
At the pole center, we are basically dividing zero by zero in view of producing a finite quantity (the pole strength).
More precisely, $A_{nm}(\omega_n) = 0$ if $m \neq n$ and else
\begin{align}
A_{nn}(\omega_n) =& \int S_n : \frac{\partial c}{\partial \omega}(\omega_n) : S_n \nonumber \\
& - \int \mathbf{u}_n \cdot \frac{\partial (\omega^2 \rho(\omega))}{\partial \omega}(\omega_n) \mathbf{u}_n .
\label{eq17}
\end{align}
In the non dispersive case, we have
\begin{align}
A_{nn}(\omega_n) &= - 2 \omega_n \int \mathbf{u}_n \cdot \rho \mathbf{u}_n ,
\label{eq18}
\end{align}
but in the viscoelastic case we have
\begin{align}
A_{nn}(\omega_n) &= - 2 \omega_n \int \mathbf{u}_n \cdot \rho \mathbf{u}_n + \imath \int S_n : \mu : S_n ,
\label{eq19}
\end{align}
with $\mu$ the phonon viscosity tensor.
Note that $A_{nm}(\omega)$ is generally complex for all frequencies, even in the non dispersive case, since the wave solution inside the PML region is complex valued.

\bigskip
Sufficiently close to the $n$-th QNM, and assuming the spectrum is separated, a single damped pole dominates the response locally and we can approximate
\begin{align}
\alpha_n(\omega) &\approx \frac{1}{\omega - \omega_n} \frac{F_n}{A_{nn}(\omega_n)} + \Sigma_n(\omega) .
\label{eq20}
\end{align}
This simple pole form is similar to the one obtained based on the resolvent method \cite{laudePRB2018}.
It does not apply, however, to the Hamiltonian, or total energy, but to the frequency response directly.
Note that when computing the frequency response along the real axis, $\omega_n \in \mathbb{C}^*$ and $\omega \in \mathbb{R}$, so that the frequency response is finite for all $\omega$.

\subsection{Modal volume and elastic Purcell effect}
\label{sec2D}

We can now define the modal volume of each elastic QNM.
Considering some point in space $r_0$, this modal volume is defined as
\begin{align}
V_n &= \frac{A_{nn}(\omega_n)}{2 \omega_n [\rho(\mathbf{r}_0) U_n^2(\mathbf{r}_0)]}
\label{eq21}
\end{align}
with the squared total displacement $U_n^2(\mathbf{r}) = u_{1n}^2(\mathbf{r}) + u_{2n}^2(\mathbf{r}) + u_{3n}^2(\mathbf{r})$.
With this definition, $V_n$ is expressed in units of cubic meters and can be thought of as measuring the volume occupied by the particular mode.
Note that the modal volume thus defined is complex-valued.
The downside of this definition is the arbitrary choice for the center position $\mathbf{r}_0$; following Ref. \cite{sauvanPRL2013}, we pick the maximum of the modal field associated with the QNM.
Specifically, since the displacements are also complex-valued, we select
\begin{align}
\mathbf{r}_0 &= \argmax_{\mathbf{r}} |\rho(\mathbf{r}) U_n^2(\mathbf{r})|.
\label{eq22}
\end{align}
A benefit of that choice is the insensitivity of the modal volume to multiplication of the modal displacement by an arbitrary complex number.
Indeed, QNMs are defined up to a complex multiplication constant only.

Furthermore, an elastic Purcell effect can be defined.
From \eqref{eq21} we have
\begin{align}
\mathbf{u}(\omega) &\approx \frac{1}{\omega - \omega_n} \frac{1}{2 \omega_n [\rho(\mathbf{r}_0) U_n^2(\mathbf{r}_0)]} \frac{F_n}{V_{n}} \mathbf{u}_n .
\label{eq23}
\end{align}
At resonance, $\omega \approx \Re \omega_n$ and $\omega - \omega_n \approx - i \Im \omega_n$.
Introducing the quality factor $Q_n = - \Re \omega_n / (2 \Im \omega_n)$, the response at resonance is then
\begin{align}
\mathbf{u}(\Re \omega_n) &\approx - i \frac{1}{\omega_n \Re \omega_n [\rho(\mathbf{r}_0) u_n^2(\mathbf{r}_0)]} \frac{Q_n}{V_n} F_n \mathbf{u}_n.
\label{eq24}
\end{align}
Numerical factors aside, the response is proportional to the Q-factor and inversely proportional to the modal volume, which are the usual signatures of Purcell's effect \cite{purcellPR1969}.
The formula is valid whatever the applied force, after projection on the QNM, so it is not limited to quantum emitters as with the original Purcell formula but it also applies to an arbitrary body force excitation.
It has been obtained here without reference to power conservation \cite{schmidtPRL2018} or an orthogonality relation \cite{elsayedPRR2020}.

\section{Applications}
\label{sec3}

\subsection{Vibrating solid rod in water}
\label{sec3A}

\begin{figure}[bt]
\centering
\includegraphics[width=80mm]{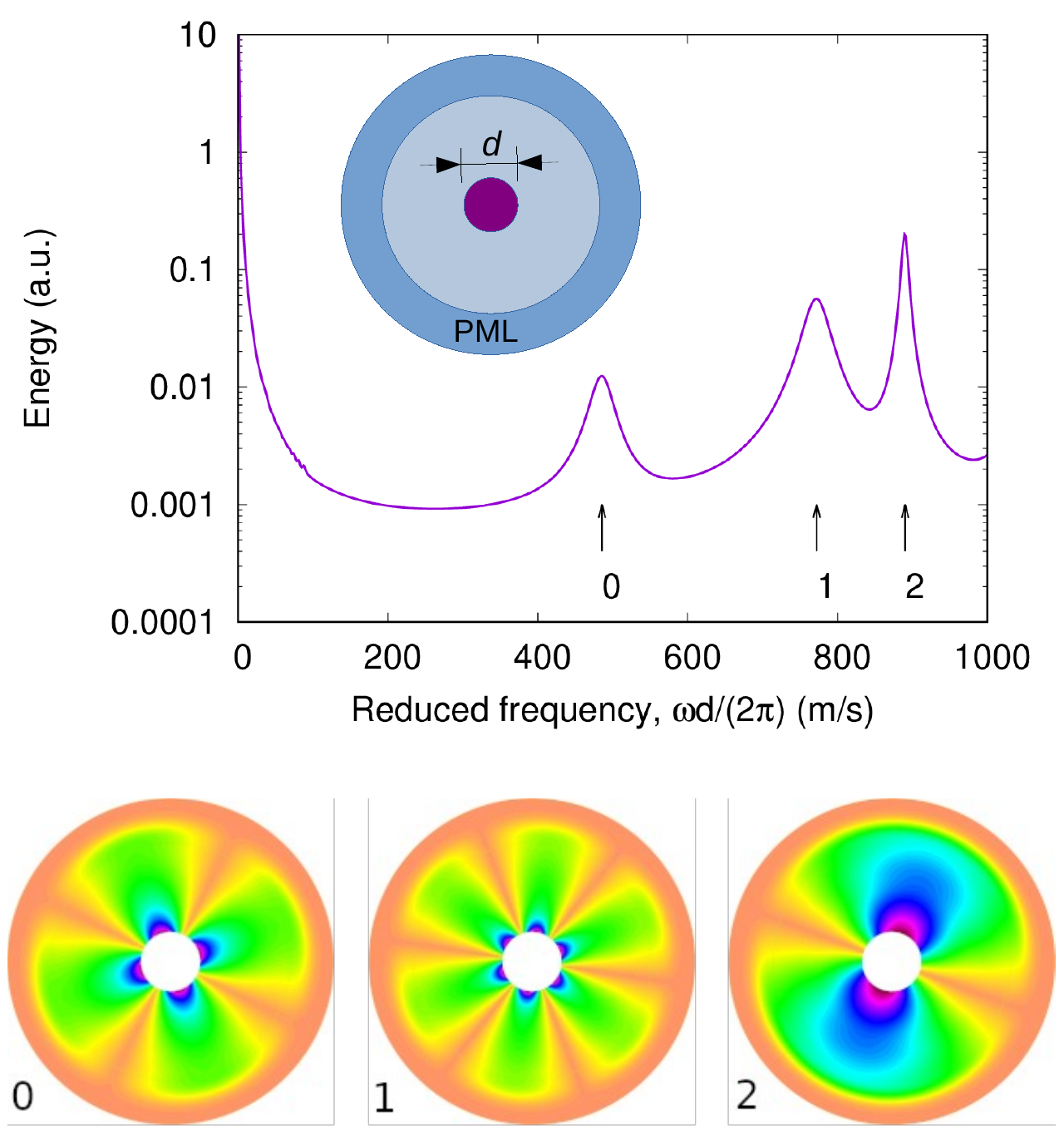}
\caption{Stochastic response for a solid nylon rod vibrating in water, as a function of the reduced frequency $\omega d/(2 \pi)$ with $d$ the diameter of the rod.
There are three damped resonances in the frequency range of interest, labeled from $0$ to $2$.
The corresponding real parts of the pressure of QNMs are shown.
The characteristics of QNMs are listed in Tab. \ref{tab1}.}
\label{fig2}
\end{figure}

As a first illustration of the concept of quasinormal mode in the context of phononics, let us consider a cylindrical rod made of nylon, immersed in water \cite{wangPRB2020}.
The elastodynamic equation is replaced in this case by a coupled acousto-elastic equation that considers the boundary conditions at the interface between the vibrating solid rod and the surrounding fluid medium in which radiation occurs \cite{wangPRB2020}.
Nylon, an isotropic solid, is chosen because the shear velocity ($1150$ \meter\per\second) is smaller than the longitudinal velocity in water ($1480$ \meter\per\second), leading to enhanced localization of elastic vibrations of the rod.
The longitudinal velocity in nylon ($2400$ \meter\per\second) is larger than in water, however.
Figure \ref{fig2}(a) shows the stochastic response \cite{laudePRB2018,wangPRB2020} of the nylon rod radiating in water.
The response is obtained by solving the acousto-elastic equation subjected to a random source distribution in the rod, as a function of frequency.
There are three damped resonances appearing in the frequency range of the plot (there are more resonances at higher frequencies).
The resonance frequencies listed in Table \ref{tab1} are well separated.
Quality factors are moderate, in the range of a few tens.
The frequency response around each peak satisfies the model of damped poles superimposed upon a background described by the \textit{ad hoc} term $\Sigma_n(\omega)$, Eq. \eqref{eq20}.

We do not attempt to describe the frequency response from the QNMs, since coupled elasto-acoustic systems are beyond the theory of Section \ref{sec2C} (see, however, Appendix B for an acoustic version of the derivation of that section), but the practical power algorithm of Section \ref{sec2B} applies straightforwardly in this case as well.
Fig. \ref{fig2}(b) shows the eigenvectors (limited to the pressure part in water) after convergence with relative error smaller then $10^{-12}$.
The obtained QNMs clearly satisfy symmetry properties that were only approximated by the maximum solutions in Ref. \cite{wangPRB2020}.

\begin{table}[tb]
\centering
\caption{Characteristics for the QNMs of a cylindrical nylon rod immersed in water. The reduced frequency is $\omega d/(2 \pi)$ with $d$ the diameter of the rod.}
\label{tab1}
\begin{tabular}{lllll}
\hline
Mode & reduced frequency (m/s) & Q \\
0 & 486 & 14 \\
1 & 772 & 20 \\
2 & 890 & 89 \\
\hline
\end{tabular}
\end{table}

\subsection{Ridge and tuning fork on a semi-infinite substrate}
\label{sec3B}

We consider next an elongated nickel ridge attached to a fused silica substrate.
The ridge is infinitely long in the third direction of space.
We thus simplify the problem of a typical elastic resonator to a two-dimensional geometry in this section, but the results would be similar in three dimensions, for instance when describing radiation from a vibrating rod \cite{benchabanePRA2017,raguinNC2019,benchabanePRA2021}.
Geometrical parameters are height $h=1000$ nm and width $w=100$ nm.
The elongated ridge has low frequency bending resonances, of the clamped-free type with in-plane polarization, but also pure shear resonances with out-of-plane displacements.
We choose this simple mechanical system because despite its simplicity it has quite well defined resonances with rather high quality factors, especially in the case of the fundamental bending vibration mode.
Because of anchoring to the silica substrate, however, the resonances are damped by radiation in the semi-infinite substrate and must hence be represented by quasinormal modes rather than normal modes.
For simplicity, material loss is not considered and elastic constants are considered non dispersive.
Notwithstanding, their inclusion per the theory of section \ref{sec2} would pose neither formal difficulty nor additional computational burden.

\begin{figure}[t]
\centering
\includegraphics[width=80mm]{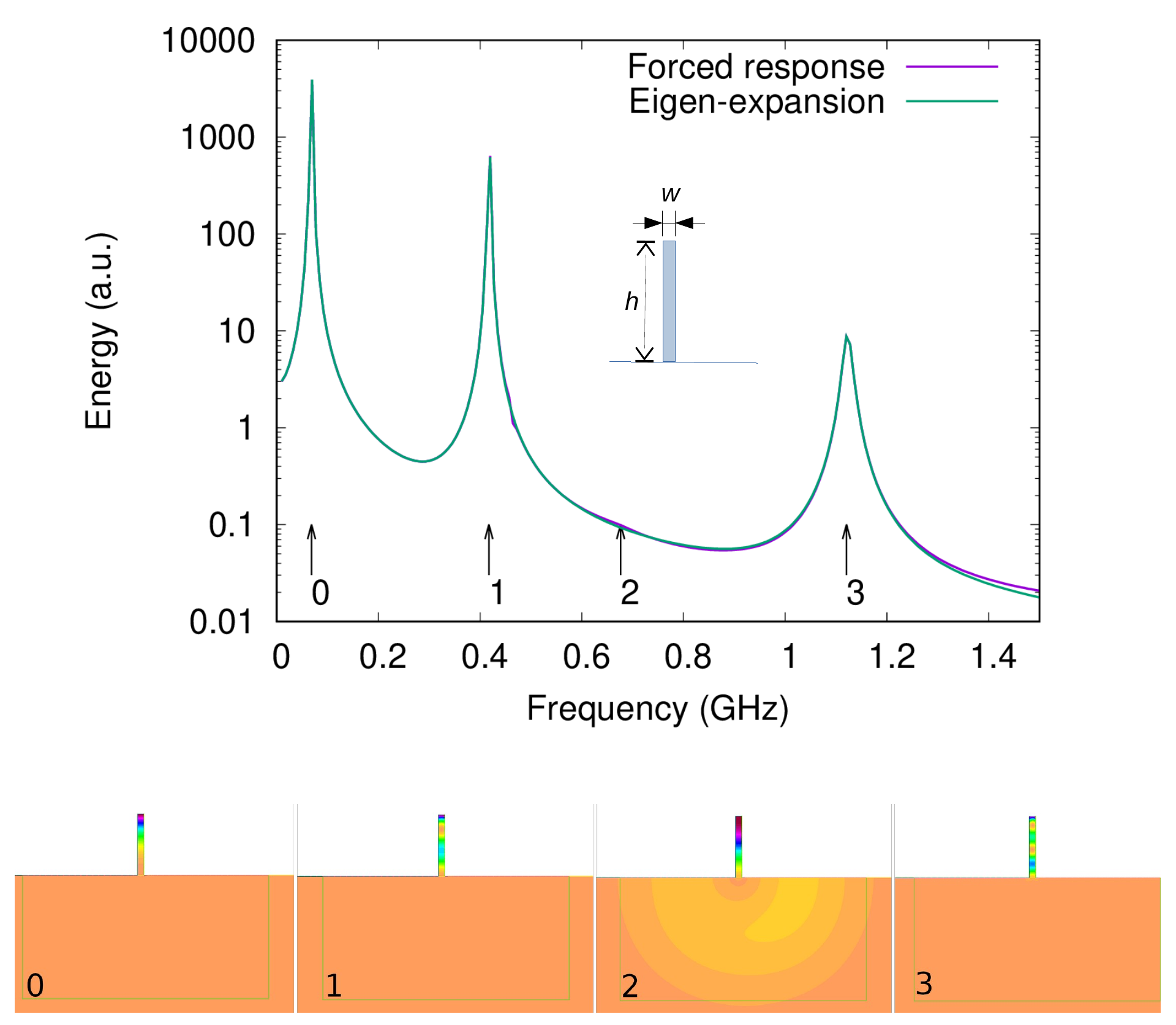}
\caption{A nickel ridge on a fused silica substrate.
The ridge is 100 \nano\meter~wide and 1000  \nano\meter~high.
(a) The frequency response is obtained for a body force on the ridge applied along the $x$-axis only ($F_x=1$).
The result of the superposition of the four quasi-normal modes identified in the frequency range of interest is plotted on top of the frequency response computed as a function of frequency by solving the forced elastodynamic equation.
(b) The modulus of the total displacement for the four QNMs is plotted.}
\label{fig3}
\end{figure}

Figure \ref{fig3}(a) shows the frequency response of the nickel ridge attached to the fused silica surface, obtained for a body force in the ridge applied along the $x$-axis.
The response shows three resonance peaks.
When we look for QNMs, we find that there are four resonances in the frequency range of interest.
These QNMs are depicted in Fig. \ref{fig3}(b) and their characteristics are summarized in table \ref{tab2}.
The third QNM is pure-SH (shear horizontal, with pure out-of-plane polarization), whereas the other three QNMs are bending modes; the third QNM hence does not contribute to the response because it is polarized orthogonal to the body force.
The quality factors of the different QNMs are quite different.
Anyway, the frequency response reconstructed by superposition of the QNMs using the eigen-expansion \eqref{eq12} reproduces quite accurately the exact computation.
Only three QNMs are sufficient in this case, since the body force is applied along the $x$-axis.
As a note the eigen-expansion computation is much faster that the full frequency response computation.
It can also be performed for any arbitrary applied body force.

The modal volumes (here expressed as the modulus of the modal in-plane area) are much smaller for the bending QNMs compared to the SH QNM.
All of them, however, are smaller than the ridge area, $0.1$ \micro\meter\squared.
There are thus all clearly confined to the surface and localized inside the ridge in the case of bending QNMs.

\begin{table}[tb]
\centering
\caption{Characteristics of the elastic QNMs of Fig.~\ref{fig3}.}
\label{tab2}
\begin{tabular}{lllll}
\hline
Mode & freq. (GHz) & Q & volume (\micro\meter\squared) & polarization \\
0 & 0.0681 & 12600 & 0.0272 & $(0.997, 0.003, 0)$ \\
1 & 0.417 & 1000  & 0.0277 & $(0.979, 0.021, 0)$ \\
2 & 0.676 & 9     & 0.0923 & $(0, 0, 1)$ \\
3 & 1.12   & 60    & 0.0285 & $(0.959, 0.041, 0)$ \\
\hline
\end{tabular}
\end{table}

\begin{figure}[t]
\centering
\includegraphics[width=80mm]{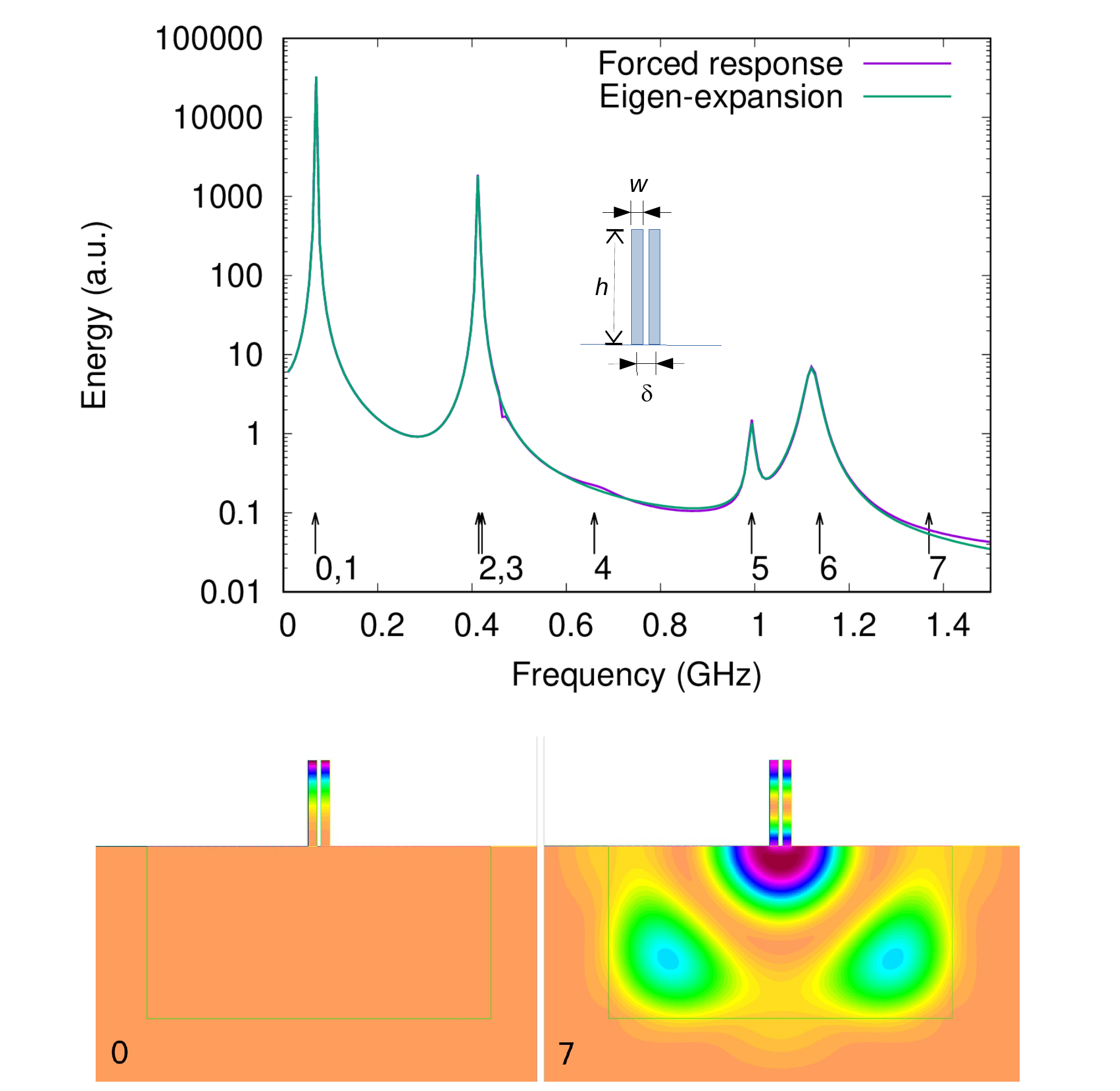}
\caption{A pair of nickel ridges on a fused silica substrate.
The ridges are $w=100$ \nano\meter~wide and $h=1000$ \nano\meter~high, and are separated by $\delta=150$ \nano\meter~from center to center.
(a) The frequency response is obtained for a body force on the left ridge only, applied along the $x$-axis only ($F_x=1$).
The result of the superposition of the eight quasi-normal modes identified in the frequency range of interest is plotted on top of the frequency response computed as a function of frequency by solving the forced elastodynamic equation.
(b) The modulus of the total displacement for two of the elastic QNMs is plotted.}
\label{fig4}
\end{figure}

Figure \ref{fig4} next considers the case of a system of two identical ridges, each identical to the one in Fig. \ref{fig3}, separated by only 50 \nano\meter ~(the center-to-center separation between ridges is $\delta=150$ \nano\meter).
As the two resonators are placed very close, they couple through the substrate and form a kind two-dimensional tuning fork.
Per the symmetry of the structure, that has a mirror plane in between the two ridges, it could be expected that each of the elastic QNMs of the single ridge doubles into a symmetric/antisymmetric pair of QNMs, similar to binding/anti-binding dimers.
It would then be expected that anti-binding QNMs show an improved quality factor and the converse conclusion for binding QNMs.
The actual situation does not follow exactly this simple intuition (see appendix C for a simple model supporting the above discussion).
The elastic QNMs characteristics summarized in Table \ref{tab3} suggest that the binding/anti-binding dimer picture applies to the first two pairs of bending QNMs (pairs $0/1$ and $2/3$ correspond resectively to QNMs 0 and 1 of the single ridge).
QNM $4$ is similar to QNM $2$ for the single ridge, with the exception of a much larger quality factor; maybe the binding QNM could not be found in this case because of a too small quality factor.
The situation is similar for QNM $6$ that is similar to QNM $3$ for the single ridge.
QNM $5$, however, has an almost pure vertical shear polarization and no counterpart in the QNMs of the single ridge.
QNM $7$ also has no counterpart in the QNMs of the single ridge.

\begin{table}[tb]
\centering
\caption{Characteristics of the elastic QNMs of Fig.~\ref{fig4}.}
\label{tab3}
\begin{tabular}{lllll}
\hline
Mode & freq. (GHz) & Q & volume (\micro\meter\squared) & polarization \\
0 & 0.0674 & 240000 & 0.0536 & $(0.997, 0.003, 0)$ \\
1 & 0.0689 & 6700   & 0.0533 & $(0.997, 0.003, 0)$ \\
2 & 0.414  & 334    & 0.0569 & $(0.979, 0.021, 0)$ \\
3 & 0.421  & 4200   & 0.0524 & $(0.972, 0.028, 0)$ \\
4 & 0.659  & 3200   & 0.113 & $(0, 0, 1)$ \\
5 & 0.993  & 82     & 0.103 & $(0.01, 0.99, 0)$ \\
6 & 1.137  & 200    & 0.0365 & $(0.957, 0.043, 0)$ \\
7 & 1.369  & 10     & 0.474 & $(0, 0, 1)$ \\
\hline
\end{tabular}
\end{table}

It is checked again that the eigen-expansion formula reproduces very closely the response to a given applied force, taken again as $F_x=1$ by applied only to the left ridge.
Only the six elastic QNMs that are polarized in the $(x,y)$ plane contribute, as for the single ridge.
In practice, the eigen-expansion response is much faster to compute than the full frequency response.
This property can be employed to compute the response to different body forces, since only the right-hand side changes with the applied force in Eq. \eqref{eq13}.

\section{Conclusion}

Elastic quasinormal modes are the eigenmodes of resonant open phononic structures subject to radiation and material loss.
As they are non conservative solution to the elastodynamic equation, their eigenfrequencies are complex numbers.
The approximation of the frequency response function to an arbitrary body force from the set of elastic QNMs appearing in the frequency range of interest was considered.
It was verified that only a small number of QNMs are required.
The derivation we have followed uses a complex, unconjugated form of the reciprocity relation for elastodynamics.
It avoids assuming energy conservation or a normalization relation and directly gives the frequency response by solving a small linear problem at each frequency.
The modal volume of elastic QNMs defined in the process is complex-valued and a formula describing the elastic Purcell's effect was obtained.
The theory extends straightforwardly to acoustic waves in open fluid media containing resonators.

\section*{Acknowledgments}

This work was supported by the EIPHI Graduate School [grant number ANR-17-EURE-0002].
Y.F.W. acknowledges support from the National Natural Science Foundation of China (Grant Nos. 12072223, 12122207 and 12021002).

\appendix

\section{Rayleigh-Lamb Reciprocity}

According to Auld (chapter 10 of his book \cite{auldBOOK1973}, starting with equation (10.106)), reciprocity for elastic waves is obtained as follows.
Equations of propagation are written for stresses and velocity as
\begin{align}
\nabla \cdot T &= \frac{\partial}{\partial t}(\rho \mathbf{v}) - \mathbf{F} , \\
\nabla \mathbf{v} &= \frac{\partial S}{\partial t} , \\
S &= s:T ,
\end{align}
with $s=c^{-1}$ the compliance tensor.
Introducing a $(\mathbf{v}, T)$ state vector with 9 components, equations are condensed as
\begin{align}
\begin{pmatrix}
0 & \nabla \cdot \\ \nabla & 0
\end{pmatrix}
\begin{pmatrix}
\mathbf{v} \\ T
\end{pmatrix} &=
\frac{\partial}{\partial t}
\begin{pmatrix}
\rho & 0 \\ 0 & s:
\end{pmatrix}
\begin{pmatrix}
\mathbf{v} \\ T
\end{pmatrix} +
\begin{pmatrix}
-\mathbf{F} \\ 0
\end{pmatrix} .
\label{eqA4}
\end{align}

Next one considers two different solutions to the equations, obtained for different forces but the same frequency (and for material constants independent of frequency).
Taking the cross-scalar product and subtracting, one obtains
\begin{align}
\nabla \cdot (\mathbf{v}_1 \cdot T_2 - \mathbf{v}_2 \cdot T_1) &= \mathbf{v}_2 \cdot \mathbf{F}_1 - \mathbf{v}_1 \cdot \mathbf{F}_2 .
\end{align}
This is a local expression, called Rayleigh or Lamb reciprocity.

A more general expression is obtained by avoiding any asumption regarding the time dependence, with two additional terms that were cancelling in the monochromatic case
\begin{align}
\nabla \cdot (\mathbf{v}_1 \cdot T_2 &- \mathbf{v}_2 \cdot T_1) = \mathbf{v}_2 \cdot \mathbf{F}_1 - \mathbf{v}_1 \cdot \mathbf{F}_2 \nonumber \\
&+ T_2 : \dot{S}_1 - T_1 : \dot{S}_2 \nonumber \\
&+ \mathbf{v}_1 \cdot \frac{\partial}{\partial t}(\rho \mathbf{v}_2) - \mathbf{v}_2 \cdot \frac{\partial}{\partial t}(\rho \mathbf{v}_1) .
\end{align}
The latter form is compatible with dispersion of the material constants.

Generally, when the integral form of this equation is obtained by integration over an infinite domain of definition, the divergence does not contribute, thanks to the radiation boundary conditions at infinity.
For waveguide problems, integration over the cross-section of the waveguide leads to an orthogonality relation for guided waves, that is a bilinear form similar to the Poynting vector appears from the divergence term.

As a remark, the expression of the reciprocity theorem is numerically inefficient, unless the solutions are known everywhere.
Solving \eqref{eqA4} is possible by the finite element method, but there are 9 equations, hence 9 unknowns compared to 3 with the displacement formulation of Section \ref{sec2C}.

\section{Sauvan’s method transposed to acoustic waves}

In this appendix, we parallel the derivation of Sec. \ref{sec2} for the case of acoustic waves in fluids.
The acoustic equation at frequency $\omega$ replacing the elastodynamic equation \eqref{eq6} is
\begin{align}
- \nabla \cdot (\rho^{-1} \nabla p) - \omega^2 B^{-1} p &= \nabla \cdot (\rho^{-1} \mathbf{F}) = g \label{eqB1}
\end{align}
for pressure field $p(\mathrm{r})$ (a scalar field) and body force $\mathbf{F}(\mathrm{r})$.
$B(\mathrm{r})$ is the elastic modulus and can be dispersive.
The scalar source field $g(\mathrm{r})$ is introduced for convenience.
Eq. \eqref{eq9} becomes
\begin{align}
\int \nabla p_2 \rho_1^{-1} \nabla p_1 - \omega_1^2 \int p_2 B_1^{-1} p_1 &= \int p_2 g_1
\label{eqB2}
\end{align}
with $\rho_1^{-1} = \rho^{-1}(\omega_1)$ and $B_1^{-1} = B^{-1}(\omega_1)$.
Eq. \eqref{eq10} is now
\begin{align}
& \int \nabla p_2 [\rho_1^{-1} - \rho_2^{-1}] \nabla p_1 \nonumber \\
& ~~~~~ - \int p_2 [\omega_1^2 B_1^{-1} - \omega_2^2 B_2^{-1} ] p_1 \nonumber \\
& ~~~~~ = \int p_2 g_1 - p_1 g_2.
\label{eqB3}
\end{align}
This is an acoustic reciprocity relation without complex conjugation, valid for an arbitrary frequency-dependent material distribution.
Note that as for elastodynamics the integration variable is not written explicitly, for compactness of expressions, but all integrals have an implied $|J| d\mathbf{r}$ factor as in Eq. \eqref{eq8}.

Next we take solution $2$ as QNM number $n$ and solution $1$ as the current solution $p$ depending on $\omega$ as a continuous parameter, such that Eq. \eqref{eq11} becomes
\begin{align}
& \int \nabla p_n [\rho^{-1}(\omega) - \rho_n^{-1}] \nabla p \nonumber \\
& ~~~~~ - \int p_n [\omega^2 B^{-1}(\omega) - \omega_n^2 B_n^{-1} ] p \nonumber \\
& ~~~~~ = \int p_n g, \forall n .
\label{eqB4}
\end{align}
The QNMs constitute a basis for the solution (per the eigenfunction expansion theorem), according to which we can write
\begin{align}
p(\omega) &= \sum_m \beta_m(\omega) p_m .
\label{eqB5}
\end{align}
Inserting the eigenfunction decomposition we obtain
\begin{align}
\label{eqB6}
\sum_m D_{nm}(\omega) \beta_m(\omega) = \int p_n g = g_n, \forall n
\end{align}
with
\begin{align}
\label{eqB7}
D_{nm}(\omega) &= \int \nabla p_n [\rho^{-1}(\omega) - \rho_n^{-1}] \nabla p_m \nonumber \\
&- \int p_n [\omega^2 B^{-1}(\omega) - \omega_n^2 B_n^{-1} ] p_m .
\end{align}
If the QNMs are known, the $D_{nm}(\omega)$ coefficients are easily computed, and the $\beta_m(\omega)$ are obtained by solving a small linear problem as a function of frequency, formally $\mathbf{\beta}(\omega) = D(\omega)^{-1} \mathbf{g}$.
It is clear that $D_{nm}(\omega_n) = 0$ by construction.
Applying the reciprocity relation \eqref{eqB2} with $\omega_1 = \omega_m$ and $\omega_2 = \omega_n$, we also have $D_{nm}(\omega_m) = 0$ for $m \neq n$.
For all other frequencies, however, $D_{nm}(\omega)$ has in principle a non vanishing value that must be taken into account in the solution.
It is then apparent that matrix $D(\omega)$ is singular at each QNM, in the complex plane, but is always invertible for $\omega$ taken along the real axis.
Finally, equation \eqref{eqB5} gives the general solution, i.e. the frequency response of the system to an arbitrary body force distribution.

If the material constants are non dispersive, the formulas simplify to
\begin{align}
D_{nm}(\omega) &= (\omega_n^2 - \omega^2) \int p_n B^{-1} p_m.
\label{eqB8}
\end{align}
Anyhow, the orthogonality relation of normal modes does not apply and matrix $D(\omega)$ is not diagonal.
The explicit expansion \eqref{eq6} still does not apply.

\bigskip
More can be said regarding the form of the solution close to a resonance, that is in the vicinity of a particular $\omega_n$.
Defining
\begin{align}
E_{nm}(\omega) &= \frac{1}{\omega - \omega_m} D_{nm}(\omega) ,
\label{eqB9}
\end{align}
$E_{nm}(\omega_n) = 0$ if $m \neq n$ and else
\begin{align}
E_{nn}(\omega_n) =& \int \nabla p_n \frac{\partial \rho^{-1}}{\partial \omega}(\omega_n) \nabla p_n \nonumber \\
& - \int p_n \cdot \frac{\partial (\omega^2 B^{-1}(\omega))}{\partial \omega}(\omega_n) p_n .
\label{eqB10}
\end{align}
In the non dispersive case, we have
\begin{align}
E_{nn}(\omega_n) &= - 2 \omega_n \int p_n B^{-1} p_n .
\label{eqB11}
\end{align}
$E_{nm}(\omega)$ is generally complex for all frequencies, even in the non dispersive case, since the wave solution inside the PML region is complex valued.

\bigskip
Sufficiently close to the $n$-th QNM, and assuming the spectrum is separated, a single damped pole dominates the response locally and we can approximate
\begin{align}
\beta_n(\omega) &\approx \frac{1}{\omega - \omega_n} \frac{g_n}{E_{nn}(\omega_n)} + \Sigma_n(\omega) .
\label{eqB13}
\end{align}
We can now define the modal volume of each acoustic QNM.
Considering some point in space $r_0$, this modal volume is defined as
\begin{align}
V_n &= \frac{E_{nn}(\omega_n)}{2 \omega_n [B^{-1}(\mathbf{r}_0) p_n^2(\mathbf{r}_0)]} .
\label{eqB14}
\end{align}
With this definition, $V_n$ is expressed in units of cubic meters and can be thought of as measuring the volume occupied by the particular mode.
Note that the modal volume thus defined is complex-valued.
The downside of this definition is the arbitrary choice for the center position $\mathbf{r}_0$; following Ref. \cite{sauvanPRL2013}, we pick the maximum of the modal field associated with the QNM.
Specifically, since the pressure is complex-valued, we select
\begin{align}
\mathbf{r}_0 &= \argmax_{\mathbf{r}} |B^{-1}(\mathbf{r}) p_n^2(\mathbf{r})|.
\label{eqB15}
\end{align}

Furthermore, an acoustic Purcell effect can be defined.
From \eqref{eqB15} we have
\begin{align}
p(\omega) &\approx \frac{1}{\omega - \omega_n} \frac{1}{2 \omega_n [B^{-1}(\mathbf{r}_0) p_n^2(\mathbf{r}_0)]} \frac{g_n}{V_{n}} p_n .
\label{eqB16}
\end{align}
At resonance, $\omega \approx \Re \omega_n$ and $\omega - \omega_n \approx - i \Im \omega_n$.
Introducing the quality factor $Q_n = - \Re \omega_n / (2 \Im \omega_n)$, the response at resonance is then
\begin{align}
p(\Re \omega_n) &\approx - i \frac{1}{\omega_n \Re \omega_n [B^{-1}(\mathbf{r}_0) p_n^2(\mathbf{r}_0)]} \frac{Q_n}{V_n} g_n p_n.
\label{eqB17}
\end{align}

\section{Radiation from single and coupled pair of ridges on a surface}

The single ridge is a vertically elongated structure, connected to the substrate only by a short segment of length $w=100$ \nano\meter.
As Figure \ref{fig3} illustrates, the displacement fields of elastic QNMs inside the ridge vary mostly along the $y$-axis but are mostly uniform along the $x$-axis.
Radiation inside the substrate then originates from a mostly uniform force distribution along the line segment (the interface between the ridge and the substrate).
Given the eigenfrequency, the wavelength inside the substrate is $\lambda = 2 \pi v / \omega$, with $v$ the phase velocity of relevant bulk elastic waves, either shear or longitudinal.
The shortest wavelength is obtained for the shear bulk wave, with velocity of 3763 m/s.
For the largest eigenfrequency of Table \ref{tab2}, $\lambda=3.36$ \micro\meter~and hence the ridge is a deep sub-wavelength structure for all considered frequencies.
As a result, radiation at infinity has the form of cylindrical bulk waves originating from the short interface line segment and is essentially monopolar.

Moving to the coupled pair of ridges separated by center-to-center distance $\delta=150$ \nano\meter, surface coupling leads to QNMs that are hybridizations of the single ridge QNM.
Owing to symmetry, hybridization leads to either binding (two small sources radiating in phase) or anti-binding (two small sources radiating in phase opposition).
The former case leads to constructive interference in the far field and hence enhanced radiation loss, whereas the latter case leads to destructive interference in the far field and hence to reduced radiation loss.
These simple considerations support the idea that anti-binding QNMs have improved Q-factors, whereas binding QNMs have deteriorated Q-factors, compared to the single ridge structure.


%

\end{document}